\documentclass[twocolumn]{revtex4}
\usepackage{graphicx}
\usepackage{dcolumn}
\usepackage{bm}
\usepackage{amsmath}
\usepackage{amsfonts}
\usepackage{amssymb}
\usepackage{color}
\usepackage{epstopdf}
\providecommand{\U}[1]{\protect\rule{.1in}{.1in}}
\newcommand{\ket}[1]{|#1\rangle}
\newcommand{\bra}[1]{\langle#1|}

\newcommand{\recto}[1]{\left[{#1}\right]}

\begin{document}
\title{Entropy production in the quantum walk}
\author{Andr\'es Vallejo$^{1,2}$}
\author{Alejandro Romanelli$^1$}
\author{Ra\'ul Donangelo$^1$}
\author{Renato Portugal$^2$}
\affiliation{\begin{small}$^1$ Facultad de Ingenier\'{\i}a, Universidad  de la Rep\'ublica, Montevideo, Uruguay\\
$^2$ Laboratorio Nacional Computa\c{c}\~ao Cient\'ifica (LNCC), Petropolis, RJ, Brazil\end{small}}
\date{\today}
\begin{abstract}
\noindent
We explore the notion of generated entropy in open quantum systems. We focus on the study of the discrete-time quantum
walk on the line, from the entropy production perspective.
We argue that the evolution of the coin can be modeled as an open two-level
system that exchanges energy with the lattice at some effective temperature
that depends on the initial state.
The entropy balance shows that there is a  positive entropy production during
the evolution, in accordance with the second law of thermodynamics.
\end{abstract}

%\pacs{03.67-a, 05.45Mt}
\maketitle

\section{\label{sec:level1}Introduction}
Over the past two decades, interest in the thermodynamic aspects of quantum systems
has increased enormously.
Very important results have been obtained, which have led to a greater
understanding of issues such as the equilibration and thermalization
mechanisms \cite{gogolin1, Linden}, the canonical typicality of quantum states
\cite{Goldstein, Popescu}, and the possibility of extracting work from quantum
systems \cite{Skrzypczyk,Quan}, among many other questions.
Some reviews on the current state of the discipline can be found in
Refs.~\cite{Gemmer,Parrondo,Goold,kossloff}.\\

One of the first questions that is natural to ask is whether the different
magnitudes of classical thermodynamics have a counterpart in quantum systems
in such a way that the laws of thermodynamics are preserved in the quantum regime.
This is not an easy question to answer, given that the inherently statistical
character of the macroscopic thermodynamic properties loses its meaning when
studying systems with only a few components.
In order to shed light on these issues, the approaches to quantum thermodynamics
from paradigms such as the quantum mechanics of open systems or from information
theory have been very helpful.

With this idea in mind, we explore the concept of generated entropy in quantum
systems.
Generated entropy is a key concept in classical thermodynamics, since it allows
to establish the degree of irreversibility associated with a thermodynamic process
or, equivalently, it can be associated with the work we could have obtained in an
ideal process between the same initial and final states, which is lost due to
irreversibilities.

For a classical system that undergoes an infinitesimal process, exchanging energy
with an environment at temperature $T$, one possible statement of the Second Law
of Thermodynamics is
\begin{equation}\label{segundaley}
dS=\frac{\delta Q}{T}+\delta S_{gen},\hspace{1cm} \delta S_{gen}\geq 0,
\end{equation}
where $dS$ is the entropy change of the system, $\delta Q$ is the heat exchanged
with the environment, $\delta S_{gen} \geq 0$ is the generated entropy
associated to the process, and the equality holds for reversible processes.

Some previous works have addressed the study of entropy production, particularly
from an information-theoretic point of view \cite{Esposito, Breuer}, and the experimental determination of entropy production rates in quantum systems has been reported recently \cite{Brunelli,Peterson}.
In this work, we analyze the validity of the expression (\ref{segundaley})
in the paradigmatic model of the discrete-time quantum walk on the line (DTQW), from a pure thermodynamic
approach.

This paper is organized as follows.  In section II, we
describe the discrete-time quantum walk on the line
(DTQW), in particular its evolution when the initial
state is a Gaussian distributed walker on the positions.
In section III the coin's degrees of freedom of the walker
are considered as a two-level open system, in contact
with a thermal bath, associated to the position's degrees
of freedom. Our main result, the entropy generation in
the quantum walk on the line is presented in section IV.
Finally, some remarks and perspectives are discussed in
section V.

\section{DISCRETE-TIME QUANTUM WALK ON THE LINE}

The DTQW on the line \cite{travaglione} evolves in the composed Hilbert space
$\mathcal{H}_{n}\otimes\mathcal{H}_{S}$,
where $\mathcal{H}_{n}$ is the position space spanned by the basis
$\lbrace\vert n\rangle\rbrace$, associated with integer positions in the line,
and  $\mathcal{H}_{S}$, the chirality space described by the basis $\lbrace\vert+\rangle,\vert -\rangle \rbrace$.
The dynamics is given by successive applications of the operator
\begin{equation}
U= \mathcal{T}(I_{n}\otimes U_{\theta}),
\end{equation}
\noindent where $U_{\theta}$ is a unitary evolution operator in two dimensions,
describing the quantum coin and parameterized by the coin
bias parameter $\theta$,
\begin{small}
\begin{equation}
U_{\theta}=\begin{pmatrix}
{\cos\theta}&{\sin\theta}\\
{\sin\theta}&{-\cos\theta}.
\end{pmatrix}
\end{equation}
\end{small}
Above, $\mathcal{T}$ is the conditional translation operator
\begin{equation}
\mathcal{T}=\sum_{n}\vert n+1\rangle \langle n\vert \otimes\vert +\rangle
\langle +\vert+
 \vert n-1\rangle \langle n\vert \otimes\vert -\rangle\langle -\vert,
\end{equation}
and $I_{n}$ is the identity operator in $\mathcal{H}_{n}$.

Any pure initial state can be expressed as
\begin{small}
\begin{equation}
\vert\psi(0)\rangle=\sum_{n}\vert n\rangle \otimes [ a_{n}(0)\vert
+\rangle+b_{n}(0)\vert -\rangle].
\end{equation}
\end{small}

\noindent where $a_{n}(0)$ and $b_{n}(0)$ satisfy the normalization condition
$\sum_{n}\vert a_{n}(0)\vert^{2}+\vert b_{n}(0)\vert^{2}=1$.
After $t$ applications
of $U$, the state will be
\begin{small}
\begin{equation}\label{estadoQW}
\vert\psi(t)\rangle=U^{t}\vert\psi (0)\rangle=\sum_{n}\vert n\rangle
\otimes [ a_{n}(t)\vert +\rangle+b_{n}(t)\vert -\rangle].
\end{equation}
\end{small}

\noindent Since in general the evolution produces entanglement between the systems, the coin will find itself in a mixed state characterized by its reduced density operator
\begin{small}
\begin{equation}
\rho_{_{S}}(t)=tr_{_{E}}\ket{\psi(t)}\bra{\psi(t)}
\end{equation}
\end{small}
\noindent whose matrix expression in the basis $\lbrace \ket{+},\ket{-}\rbrace$ is
\begin{small}
\begin{equation}\label{rhocoin(t)}
\rho_{_{S}}(t)=\begin{pmatrix}
{\sum_{n}\vert a_{n}(t)\vert^{2}}&\vspace{0.2cm}{\sum_{n}a_{n}(t)b_{n}^{*}(t)}\\
{\sum_{n}a_{n}^{*}(t)b_{n}(t)}&{\sum_{n}\vert b_{n}(t)\vert^{2}}
\end{pmatrix} .
\end{equation}
\end{small}

It is well known that for large $t$, the expression (\ref{rhocoin(t)})
becomes stationary, so the coin reaches an equilibrium state that depends
on the initial state of the global system
\cite{Nayak, Carneiro, Abal, Annabestani, Omar, Pathak, Petulante, Venegas}.
In what follows we will focus on the coin evolution, considering it as a
two-level open system that equilibrates due to interaction with a large
environment, composed by the infinite position degrees of freedom of the walker.
In order to highlight the interpretation of $\mathcal{H}_{n}$ as a thermal bath,
we will consider situations in which many sites in the position Hilbert space
are initially occupied.
In the case of bipartite systems with a global Hamiltonian without degenerate energy gaps, this restriction has been proven to be sufficient to assure the equilibration on average of the system \cite{Linden}. 
We must point out that in the case of the DTQW on an infinite line, an equilibrium
state is reached even if the walker is initially localized.
However, in spite of this singular behavior, as a rule, some kind of thermodynamic
behavior can only be expected if many sites of the environment are occupied.
Therefore, we will consider this case, which on the other hand, is the
one of interest in quantum computation algorithms \cite{Renato,Grover}.
In particular, we shall consider the family of initial states
\begin{small}
\begin{equation}\label{inistateQW}
a_{n}(0)=\frac{e^{\frac{-n^{2}}{4\sigma^{2}}}}{\sqrt[4]{2\pi\sigma^{2}}}
\cos(\gamma/2),\;b_{n}(0)=
\frac{e^{\frac{-n^{2}}{4\sigma^{2}}}}{\sqrt[4]{2\pi\sigma^{2}}}\sin(\gamma/2)e^{i\varphi}
\end{equation}
\end{small}

\noindent corresponding to an initially Gaussian walker, distributed around the origin with a width $\sigma$ and  with arbitrary chirality determined by the angles $\gamma$ and $\varphi$.
The asymptotic reduced density matrix for the initial state given by Eq.(\ref{inistateQW})
has been obtained in recent works, in the limit $\sigma\gg 1$ \cite{Vallejo, Orthey}:
\begin{small}
\begin{equation}\label{coinRDM}
\overline{\rho}_{_{S}}=\dfrac{1}{2}\begin{pmatrix}
1+\cos\alpha \cos\theta &\cos\alpha \sin\theta\\
\cos\alpha \sin\theta& 1-\cos\alpha \cos\theta
\end{pmatrix} .
\end{equation}
\end{small}

\noindent where
\begin{equation}\label{alpha}
\cos\alpha=\cos\theta\cos\gamma+\sin\theta\sin\gamma\cos\varphi
\end{equation}
is the cosine of the angle between the initial Bloch vector
\begin{equation}
\vec{B}=(\sin\gamma\cos\varphi, \sin\gamma\sin\varphi,\cos\gamma)
\end{equation}
and the vector $\vec{v}$
\begin{equation}
\vec{v}=(\sin\theta,0,\cos\theta).
\end{equation}

\section{THE COIN AS AN OPEN SYSTEM}
In order to implement an entropy balance, we must analyze the DTQW on the line,
which is essentially a mathematical model, from a physical point of view.
As it was previously mentioned, we will consider the coin's degrees of
freedom as a two level system in thermal contact with a large bath,
described by the position's Hilbert space $\mathcal{H}_{n}$.
To complete the physical description, we must identify the local Hamiltonian
whose expectation value will be defined as the internal energy, and the
temperature $T$ of the lattice experienced by the qubit.
In Ref. \cite{Vallejo} it is shown that for sufficiently wide position distributions the dependence on the initial
state in Eq. (\ref{coinRDM}) can be factorized in such a way that the
asymptotic reduced density matrix $\overline{\rho}_{_{S}}$ can be written in the canonical distribution form
\begin{equation}\label{qwthermal}
\overline{\rho}_{_{S}}=\frac{e^{-\beta H_{S}'}}{tr(e^{-\beta H_{S}'})},
\end{equation}
for a fixed Hermitian operator $H_{S}'$, called entanglement Hamiltonian,
\begin{equation}\label{HentQW}
H_{S}'= -\varepsilon\vec{\sigma}.\vec{v}.
\end{equation}
Here $\varepsilon$ is an arbitrary factor with units of energy and $\vec{sigma}$ the vector whose components are the Pauli matrices.

The initial-state-dependent entanglement temperature
$T_{ent}=1/(k_{_{B}}\beta)$ is \cite{Romanelli1}:
\begin{equation}
T_{ent}=\frac{\varepsilon}{k_{B}\ln [\tan(\alpha/2)]};
\end{equation}
where $k_{B}$ is the Boltzmann constant. In addition, $T_{ent}$ contains all the dependence on the initial state, and it is a measure
of the entanglement produced during the evolution.

\begin{figure}%[H!]
 \centering
  \includegraphics[trim= 300 80 -50 0, scale=0.5, clip]{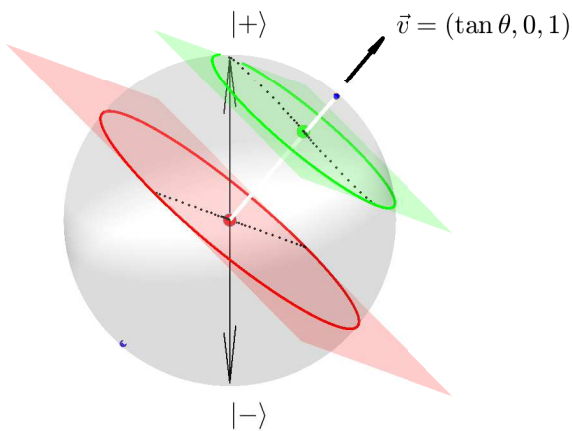}
\caption{Time evolution of the Bloch vector associated with the coin state.
The Bloch vector associated the the successive states is represented by the
black dots, that converge to the asymptotic state (the center of each circle).
The initial states considered are in both cases a Gaussian distributed walker with $\sigma=10$ and initial chirality $\vert +\rangle$ (green circle) and
$\frac{1}{\sqrt{2}}(\vert +\rangle +\vert -\rangle)$ (red circle), respectively.
The same asymptotic state is reached for all the initial coin states in each circumference.}
\label{nivelQW}
\end{figure}

We observe that the asymptotic reduced state Eq.(\ref{qwthermal}) can be interpreted as the
equilibrium state of a two-level system governed by an effective local
Hamiltonian $H_{S}'$ that equilibrates due to thermal contact with a heat
reservoir at temperature $T_{ent}$. In this case the temperature is not an intrinsic
property of the bath, since it also depends on the initial state of the system
through the parameter $\alpha$.
As a consequence, there is no global attractor for all initial states, but a
diameter of equilibrium states, one for each value of $\alpha$.
This implies that all initial states placed on a circumference orthogonal
to the vector $\vec{v}$ on the Bloch sphere, evolve to a common equilibrium state, located at the center of the circumference (see Fig.\ref{nivelQW}).

We should remark that when narrow initial position distributions are considered, the system cannot be modeled under the system-thermal bath paradigm since the equilibrium state cannot be expressed in the form of Eq. (\ref{qwthermal}) for a ﬁxed Hamiltonian \cite{Vallejo}. 
An additional argument that reinforces the idea of taking $H_{S}'$ as the local
Hamiltonian arises from the analysis of the numerical simulations.
Figure \ref{nivelQW} shows the evolution of the reduced state in the Bloch
sphere for a Hadamard walk ($\theta =\pi /4$).
We notice that for both initial states considered, the Bloch vector at the
time $t+1$ is very close to the vector that we would obtain by rotating
the Bloch vector at time $t$ an angle $\eta=\pi$ with respect to the
$\hat{e}=\frac{1}{\sqrt{2}}(1,0,1)$ axis.
This suggests that the unitary part of the reduced evolution for one step must
coincide with such rotation.
Recalling the general expression for a rotation operator,
\begin{equation}
R_{\eta,\hat{e}}=e^{i\gamma}e^{-i\frac{\eta}{2}\hat{e}.\vec{\sigma}},
\end{equation}
the unitary part of the coin evolution for one step can be obtained by choosing
$\hat{e}=\frac{1}{\sqrt{2}}(1,0,1)$, $\eta=-\pi$ and arbitrary $\gamma$
(in what follows we take $\gamma=0$ for simplicity),
\begin{equation}
U'=R_{\pi,\frac{1}{\sqrt{2}}(1,0,1)}=e^{i\frac{\pi}{2}H},
\end{equation}
where the generator of the rotation $H$ is the Hadamard operator

\begin{equation}
H=\frac{1}{\sqrt{2}}
\begin{pmatrix}
{1}&{1}\\
{1}&{-1}
\end{pmatrix} .
\end{equation}
On the other hand, the entanglement Hamiltonian, Eq. (\ref{HentQW}), for $\theta=\pi/4$, is also essentially the Hadamard operator
\begin{equation}
\label{hadamard ent_ham}
H_{S}'=-\varepsilon\vec{\sigma}.\vec{v}=-\varepsilon H .
\end{equation}
which shows that the unitary part of the observed reduced dynamics is generated
by the same operator that appears in the expression of the thermal state,
Eq.(\ref{qwthermal}), and supports our argument about considering
$H_{S}'$ to serve as the local Hamiltonian.

\section{GENERATED ENTROPY}
Once we have identified the local Hamiltonian and the effective temperature of
the thermal bath, we are in a position to implement an entropy balance. 
We will assume that the entropy of the coin corresponds to the
von Neumann entropy of its reduced state $\rho_{_{S}}$:
\begin{equation}
S_{vN}=-k_{_{B}}tr(\rho_{_{S}}\ln\rho_{_{S}}),
\end{equation}
whose expression in terms of the eigenvalues of $\rho_{_{S}}$, $\lambda_{+/-}$, is \cite{Nielsen}:
\begin{equation}\label{SvN}
S_{vN}=-k_{_{B}}[\lambda_{+}\log(\lambda_{+})+\lambda_{-}\log(\lambda_{-})]
\end{equation} 

Defining the internal energy as the expected value of the local Hamiltonian
\begin{equation} E (t)=\langle H_{S}'\rangle =-\varepsilon tr [\rho_{_{S}}(t)H],
\end{equation}
and also defining, since there is no work involved, the heat exchanged in one
step as the change in the internal energy ,
\begin{equation}
\delta Q_{t\rightarrow t+1}= E(t+1)-E(t) ,
\end{equation}
we will investigate the validity of the expression (\ref{segundaley})
of the second law of thermodynamics.
The entropy generated until step $t$, starting from a pure state,
which therefore has $S_{vN}(0)=0$, is
\begin{small}
\begin{equation}\label{secondlawQW}
S_{gen}(t)=S_{vN}(t)-\frac{Q_{0\rightarrow t}}{T_{ent}} ,
\end{equation}
\end{small}
where $Q_{0\rightarrow t}=E(t)-E(0)$ is the total heat exchanged during the
evolution.

Analyzing the numerical simulations of Fig. (\ref{nivelQW}), we notice that
the trajectory defined by the successive non equilibrium states lies very close to
the plane $\alpha =const.$
This implies that the scalar product of the Bloch vector with the vector
$\vec{v}$ that defines the Hamiltonian is approximately constant, a
fact that, because of the previous definitions, must be interpreted as the
approximate conservation of the local energy during the evolution. However, these small energy fluctuations contribute to the generations of entropy in the process, as we will show in the examples below. 

%In particular, since the initial and equilibrium energies coincide, the entropy produced until the thermal state is reached coincide with the asymptotic value of the von Neumann entropy. Using the eigenvalues of (\ref{coinRDM}), $\lambda_{\pm}=1/2\pm\cos{\alpha}/2$, it is straightforward to show that:
%\begin{equation}
%S_{gen}^{\infty}=S_{vN}(\infty)=k_{B}\ln\recto{\frac{2\recto{\tan{(\alpha /2)}}^{\cos{\alpha}}}{\sin{\alpha}}}.
%\end{equation}

\begin{figure}
 \centering
%  \subfloat[]
  {\label{f:entropygauss}
    \includegraphics[width=0.9\columnwidth]{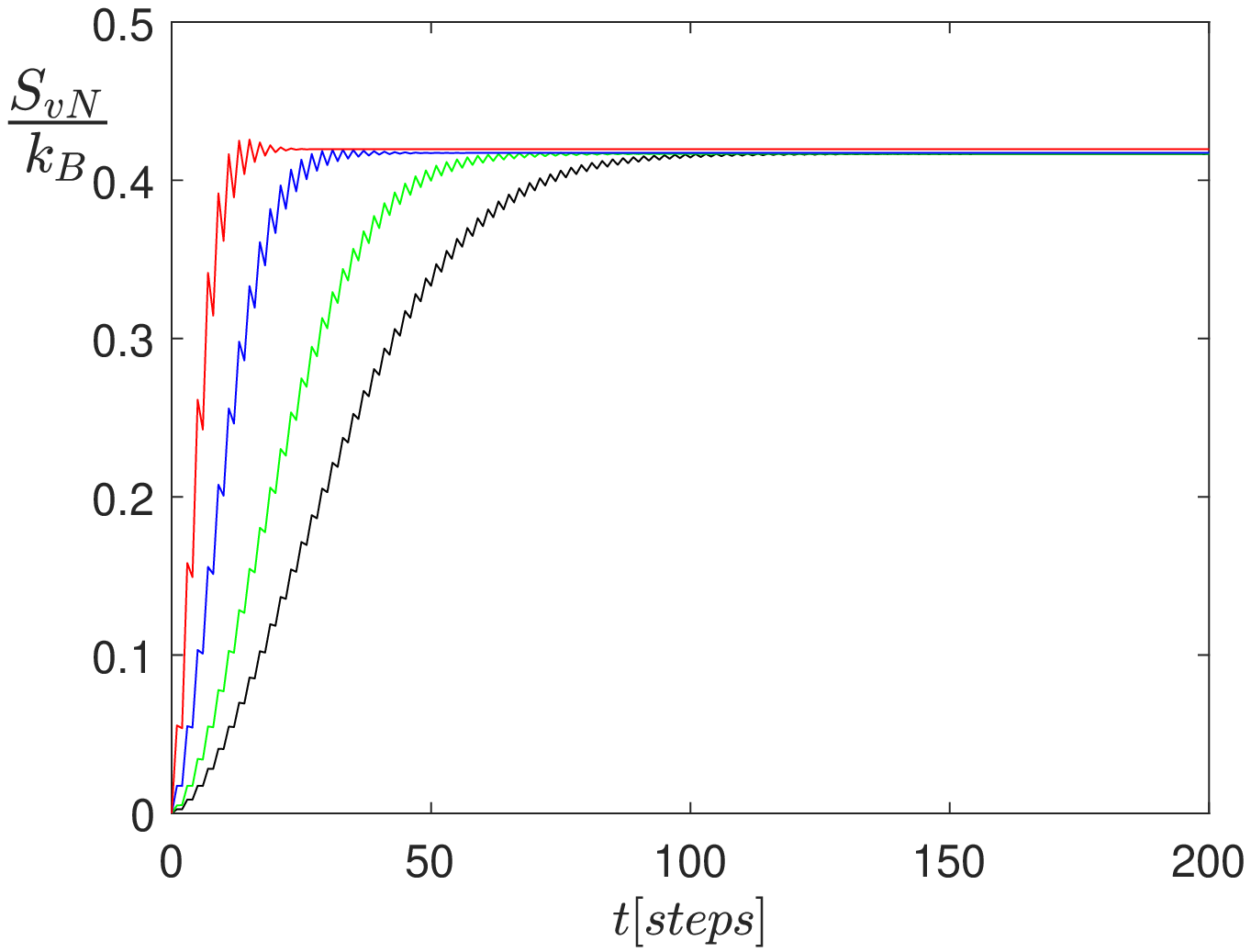}}
         (a)
%  \subfloat[]
  {\label{f:heatgauss}
    \includegraphics[width=0.9\columnwidth]{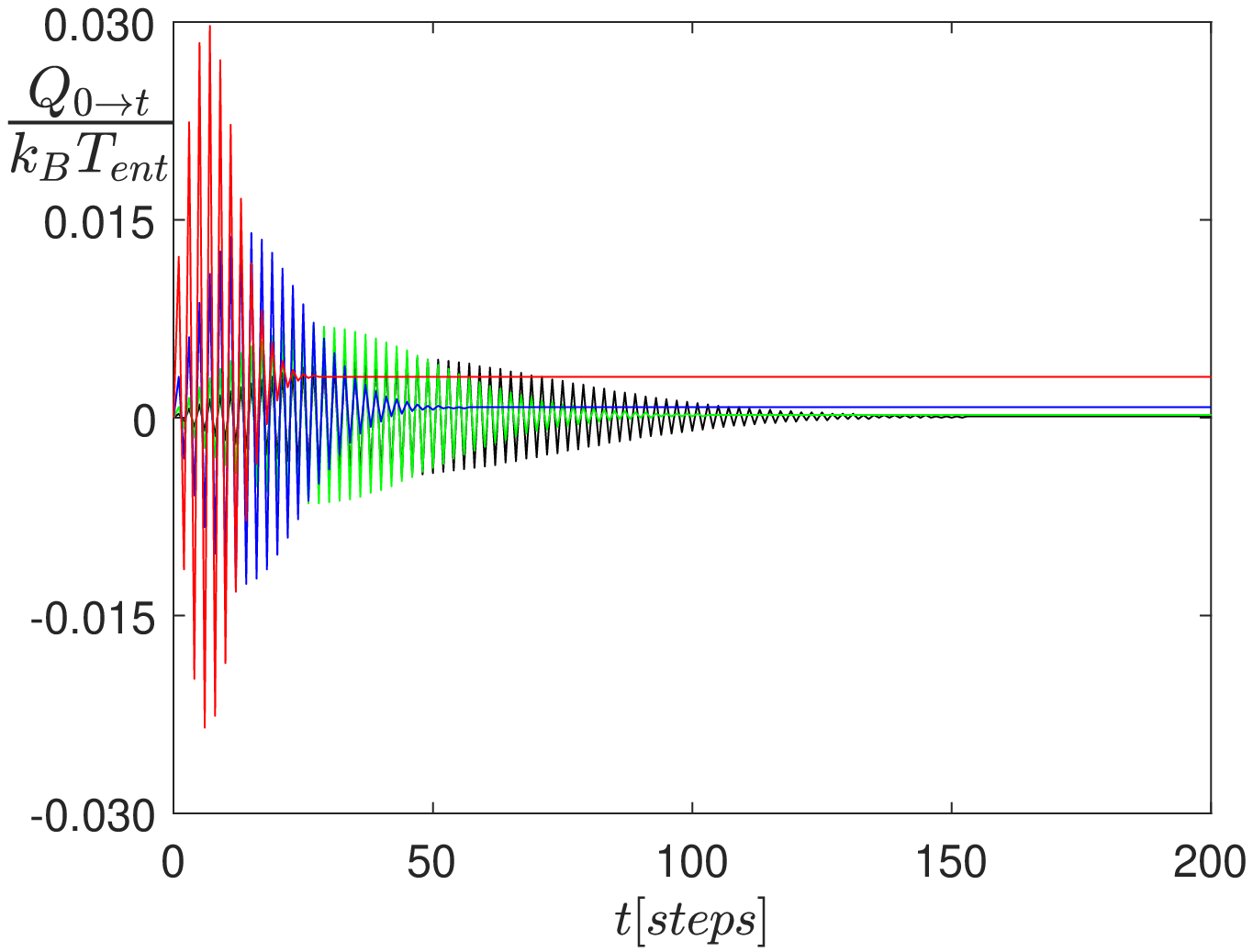}}
          (b)
% \subfloat[]
  {\label{f:sgengauss}
    \includegraphics[width=0.9\columnwidth]{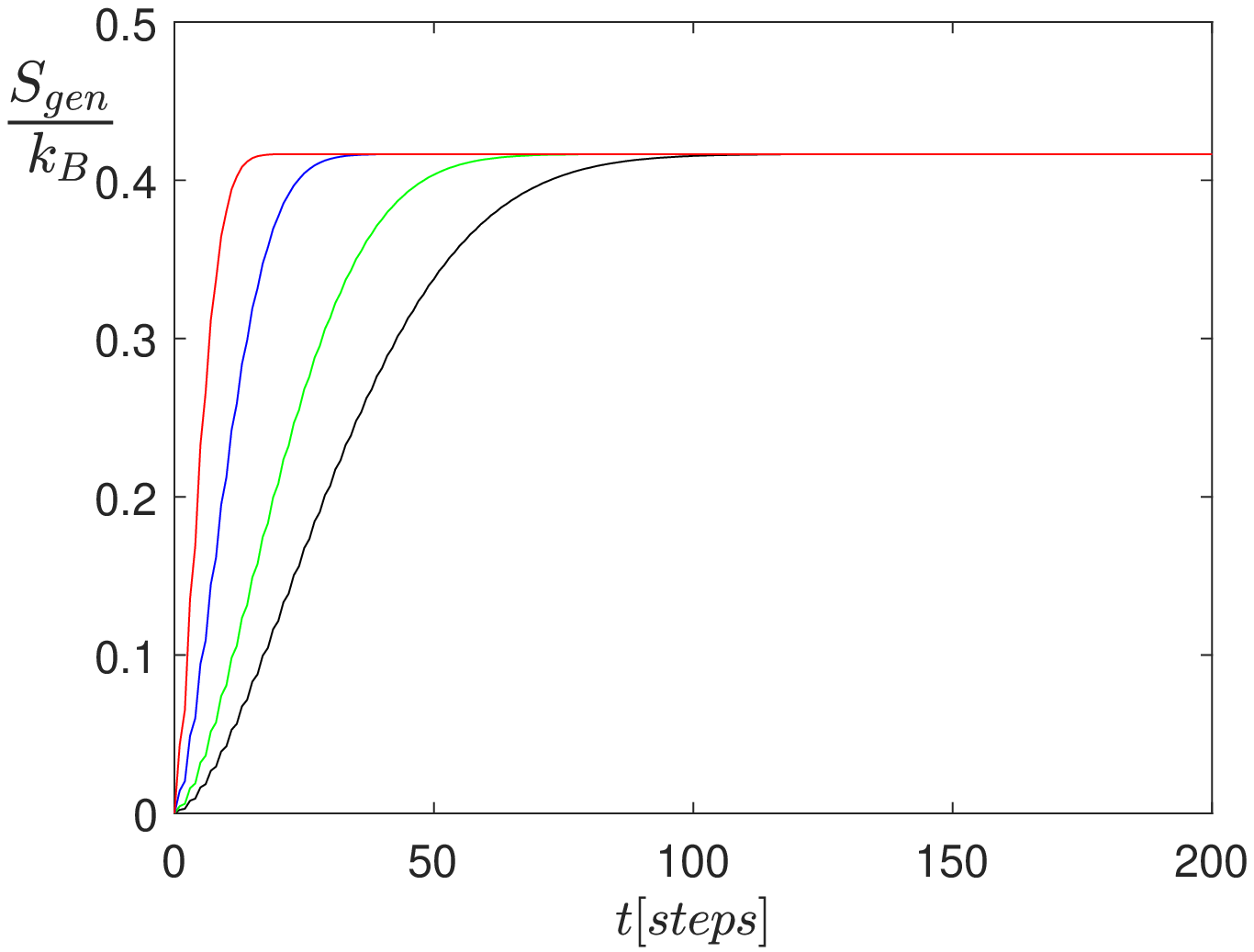}}
          (c)
 \caption{Dimensionless thermodynamic function for (a) von Neumann entropy,
    (b) entropy change due to heat transfer, and
    (c) generated entropy, corresponding to an initially Gaussian-distributed walker
    with $\sigma=30$ (black thick line), 20 (green thin line), 10 (blue dot-dashed line), and 5 (red dotted line). The initial chirality is $\vert +\rangle $}
 \label{f:Gaussian}
\end{figure}

In Fig.(2) we present the numerical results for the
von Neumann entropy as a function of the number of steps, for a Gaussian walker
initially centered at the origin and several values of $\sigma$, with the coin
initially in the state $\vert+\rangle$.
We notice that, in spite of the overall growing trend, the von Neumann entropy
oscillates  before reaching the asymptotic state.
After considering the transport term due to heat flow shown in
Fig.2(b), we notice that the generated entropy,
Fig.2(c), presents a  monotonically increasing behavior.
Therefore, whenever the von Neumann entropy of the system decreases, heat is transferred to the environment in an amount such that its increase in entropy exceeds the reduction in the von Neumann entropy.

We note that, although the final value of the entropy does not depend on $\sigma$, as long as the width does not take very small values, the evolution to the asymptotic value is faster for small values of $\sigma$. This can be understood when the evolution of the occupancy distribution is considered. In Ref. \cite{Zhang} it is shown that the initial Gaussian distribution gradually separates into two Gaussian peaks that move, to the right and to the left, with a velocity that is determined by the parameters of the evolution operator. Therefore, the time required for the separation to be complete (within a given approximation) is proportional to the width of the original position distribution. We have verified that when the separation is complete, the entropy stops changing. This explains the slower growth of $S_{gen}(t)$ for the larger values of $\sigma$. 

Numerical simulations show analogous behavior for other nonlocal initial
states of the walker, given that many states of the position space are
initially occupied.
For example, in Fig.(3) we show simulations for the case where
the initial state is an uniform superposition of several kets
$\vert n\rangle$ centered at the origin.
We note that the thermal state is in good agreement with the one obtained in the previous,
Gaussian-distributed case.
This supports the use of the same entanglement Hamiltonian in the entropy
balance.
In Fig.(3) we again notice that the consideration of the entropy
flux term corrects the von Neumann entropy oscillations in the evolution
towards the asymptotic state.

On the other hand, localized initial states do not lead to an equilibrium state that can be
written in the form of Eq. (\ref{qwthermal}), so the use of Eq. (\ref{HentQW})
in the entropy balance is not justified.
%\textbf{On the other hand, localized initial states of the walker present a different behavior. Although they also evolve towards an equilibrium state in an infinite line, the equilibrium state can not be expressed in the form of Eq. (\ref{qwthermal}), unless we allow different Hamiltonians for different initial chiralities, as it was shown in \cite{Vallejo}. Such an initial-state-dependent Hamiltonian would lead to an internal energy without the property of being a state function, since it would also depend on the initial state. From another point of view, localized initial states would imply that the position's Hilbert space can not be considered as a thermal bath, since the equilibrium state do not adopt the form of a thermal state. The impossibility of determining the Hamiltonian looking at the equilibrium state prevents the possibility of implementing the entropy balance by the present methods.

\begin{figure}[!h]

 \centering
  {\label{f:entropyuniform}
    \includegraphics[width=0.9\columnwidth]{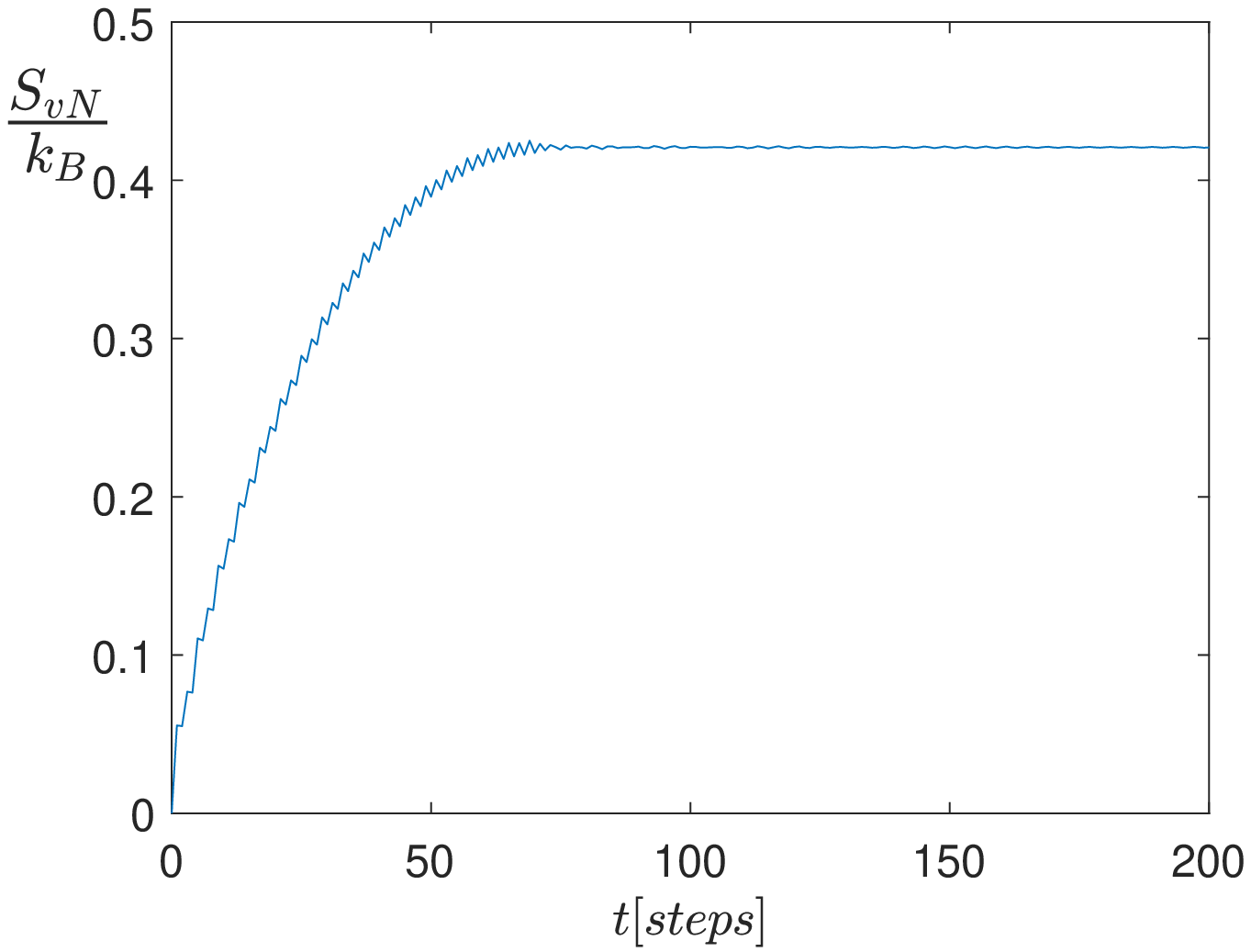}}
        (a)
%    \subfloat[a]
  {\label{f:heatuniform}
    \includegraphics[width=0.9\columnwidth]{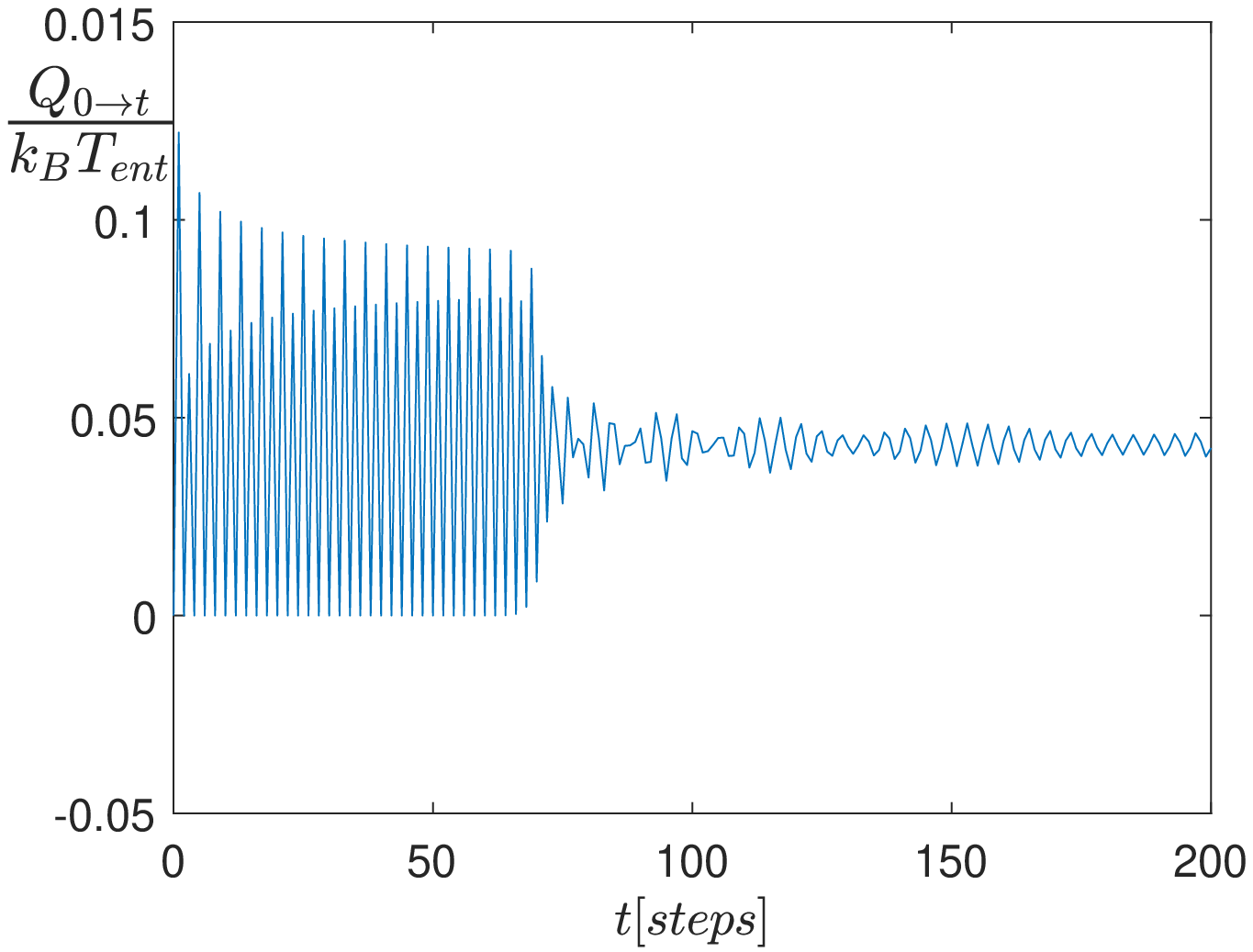}}
         (b)
%\subfloat[b]
 {\label{f:sgenuniform}
    \includegraphics[width=0.9\columnwidth]{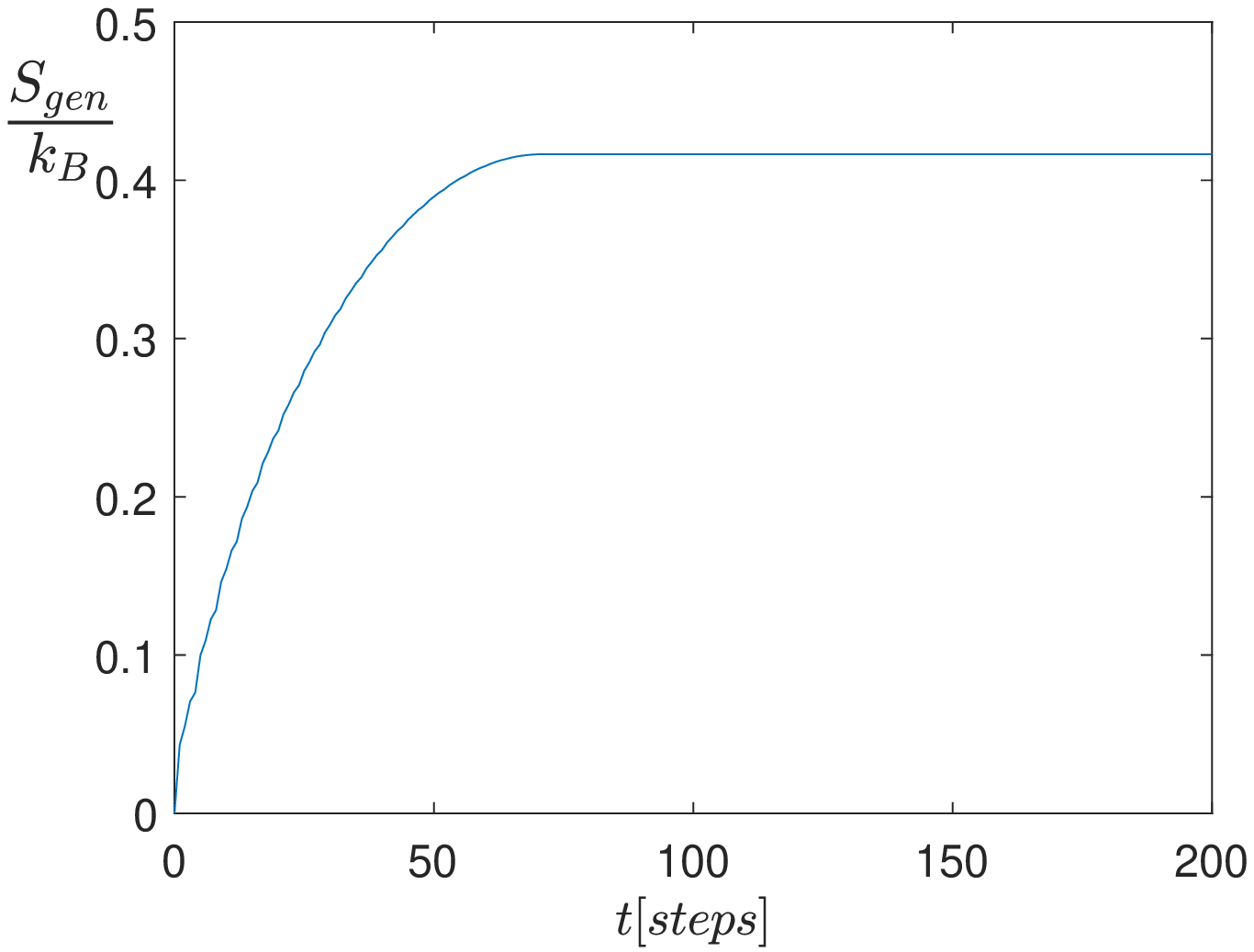}}
           (c)
% \subfloat[c]
 \caption{Similar to Fig (2) for a walker that starts in a uniform superposition of 101 position states centered at the origin. The initial chirality state is $\vert +\rangle $.The dimensionless thermodynamic function is shown for (a) von Neumann entropy.
    (b) entropy change due to heat transfer.
    (c) generated entropy}
\label{Uniform}
\end{figure}

\subsection{Consistency with previous work}
Reference \cite{Breuer} calculates the entropy generation for open Hamiltonian
systems in thermal contact with a large reservoir:
\begin{equation}\label{sgenrelent}
S_{gen}(t)=k_{B} \left[ D(\rho_{_{S}}(0)\parallel\rho_{_{S}}^{eq})
-D(\rho_{_{S}}(t)\parallel\rho_{_{S}}^{eq})\right] ,
\end{equation}
where $D(\rho\parallel\rho ')=tr(\rho\ln\rho)-tr(\rho\ln\rho ')$ is the relative
entropy of the states $\rho$ and $\rho '$, and $\rho_{_{S}}^{eq}$ is the
equilibrium state.
It is possible to show that the expression (\ref{sgenrelent})
is always positive if $\rho_{_{S}}^{eq}$ is a stationary solution of
the reduced dynamics.
This excludes systems presenting recurrences, but holds for a DTQW evolving
on an infinite line.

The consistency between our formalism and Eq.(\ref{sgenrelent}) is a direct
consequence of the possibility, for initial wide Gaussian position distributions, of writing the equilibrium state in the form of
Eq.(\ref{qwthermal}).
As an example, note that the asymptotic value of the generated entropy,
according to Eq. (\ref{sgenrelent}), is
\begin{equation}
S_{gen}^{\infty}=k_{B}D(\rho_{_{S}}(0)\parallel \rho_{_{S}}^{eq}) .
\end{equation}
Using the definition of relative entropy and the equilibrium state (\ref{qwthermal}), after some algebra we obtain
\begin{equation}\label{sgenmax}
S_{gen}^{\infty}=k_{B}\ln\recto{\frac{2\recto{\tan{(\alpha /2)}}^{\cos{\alpha}}}{\sin{\alpha}}} .
\end{equation}
\noindent This coincides with the calculation of the von Neumann entropy using the eigenvalues of Eq. (\ref{qwthermal}), as expected since in this case the heat exchanged with the environment $Q_{0\rightarrow t}$ is zero, in agrement with the discussion presented in the paragraph followig Eq. (\ref{secondlawQW}).

For a Hadamard walk with the coin starting in the state
$\ket{+}$, i.e. $\gamma =0$, and a Gaussian-distributed walker,
Eq. (\ref{alpha}) implies that $\alpha =\pi /4$.
Substituting this value in Eq. (\ref{sgenmax}), we obtain:
\begin{equation}
\frac{S_{gen}^{\infty}}{k_{B}}=\frac{3}{2}-\frac{\sqrt{2}}{2}\ln(\sqrt{2}+1)\simeq 0.4165,
\end{equation}
which coincides with the asymptotic value of Figs.2(c) and 3(c).

\section{Final remarks and conclusions}
The main objective of this work has been the study
of the DTQW on the line from the point of view of entropy generation. We considered the chirality degrees
of freedom as a two-level system that evolves towards
an equilibrium state due to its interaction with a much
larger environment, composed by the position degrees of
freedom.

It is important to emphasize that the interpretation of
$\mathcal{H}_{n}$ as a thermal bath could only be established when the
initial position occupation level is high. This means that
our present study does not include highly localized initial
states.

After identifying the local Hamiltonian, we have observed variations in the evolution of its expected value,
a fact that can be interpreted as the equivalent of heat
transfer with the lattice. This implies, for example, that
in optical implementations of quantum walks \cite{Dur, Schreiber, Rhode,Sansoni}, there
should always exist energy transfer between the photon
and the optical devices. The consideration of the heat
transfer term in the entropy balance is of particular importance, since it ensures a monotonous increase in the
entropy production during the entire evolution. Since the
time-reversed process would imply entropy destruction,
our study suggests that, despite being possible due to
unitary reversibility, the Gaussian packet narrowing in
position space, starting from a highly distributed state is
an extremely unlikely process from the thermodynamic
point of view. This is equivalent to saying that, at least
in the case of this system, the process of going from a
very entangled state to a product state is extremely improbable.

%Extensions of the present work that include the DTQW on the line with more
%coin states, for higher dimensional lattices, as well as on a finite ring,
%are currently under investigation.

\section*{Acknowledgments}
This work was partially supported by CAP, ANII, PEDECIBA (Uruguay) and CAPES (Brazil).

\end{document}